\documentclass[twocolumn,showpacs,preprintnumbers,amsmath,amssymb,english]{revtex4}

   \usepackage{graphicx}
\usepackage{dcolumn}
\usepackage{bm}

\begin{document}

\title
{Light-hole quantization in the optical response of ultrawide GaAs/Al$_x$Ga$_{1-x}$As quantum wells \\}

\author{V.V. Solovyev,$^{1,2}$, V.A. Bunakov, $^1$ S. Schmult$^{2}$ and I. V. Kukushkin$^{1,2}$}

\affiliation{
\it $^1$Institute of Solid State Physics, Russian Academy of Sciences, Chernogolovka, 142432 Russia\\
\it $^2$Max-Planck-Institute for Solid State Research, 70569 Stuttgart, Germany\\
}
\date{\today}

\begin{abstract}

Temperature-dependent reflectivity and photoluminescence spectra are studied for undoped ultra-wide 150 nm and 250 nm GaAs quantum wells. It is shown that spectral features previously attributed to a size quantization of exciton motion in z-direction coincide well with energies of quantized levels for light holes. Furthermore, optical spectra reveal very similar properties at temperatures above exciton dissociation point.

\end{abstract}

\maketitle
\smallskip

\medskip

The optical response of a semiconductor material close to the band gap energy is determined at low temperatures by excitons, the particles composed of a conduction band electron bound to a valence band hole by the Coulomb interaction. Since their discovery ~\cite{Gross} excitons provide an extremely useful model system for many-body problems of condensed matter physics. The two-dimensional nature of excitons formed in semiconductor heterostructures usually simplifies theoretical considerations if compared to the 3D case and at the same time opens up new experimental possibilities. Recently, several attempts ~\cite{Kochereshko} ~\cite{Kochereshko2} were made to study properties of excitons with a quantized energy spectrum due to a one dimensional confinement in an ultra-wide quantum well (QW). If the QW width $d$ is much larger than the Bohr radius of an exciton, a quantization procedure should consider the latter as a single particle with a total mass equal to the sum of electron and hole masses. (This model is opposed to the case of narrow QW exciton where one starts from the quantized spectra of an individual electron and a hole and then adds the Coulomb attraction between them.) The authors in ~\cite{Kochereshko} associated features in the reflectivity spectra with a set of energies on the heavy-hole exciton dispersion relation selected by a quantization in momentum space according to $K_z=N\frac{\pi}{d}$, where $N$ is a positive integer.

In this report we present a study of ultra-wide undoped GaAs quantum wells by reflectivity and photoluminescence (PL) spectroscopy. The results demonstrate that the observed spectra cannot be described by a model of size quantized excitons. Instead, the extracted energy levels fit much better into the model of light-hole quantization.

The high quality samples were grown by a molecular beam epitaxy (MBE) and comprised a slab of undoped GaAs (150 or 250 nm in width) surrounded by 200 nm and 400 nm undoped Al$_{0.32}$Ga$_{0.68}$As barriers. In PL experiments the laser excitation at 780 nm and the collected signal were guided via two corresponding optical fibers. The same setup was used for recording reflectivity spectra. The light source for the latter experiment was a filament lamp with a white light passed through a 780 nm Raman edge optical filter to suppress excitation of PL lines in detected spectra.

In Figure 1 we present photoluminescence and reflectivity spectra for both samples recorded at a low temperature of 1.5 K. A multitude of narrow (~0.2 meV in width) spectral lines are resolved in reflections with dispersive-like shapes for 150 nm QW and inverse absorptive-like shapes for the other sample. This lineshape change results from the optical path differences in the sample layers ~\cite{Lineshape}. The energy separation between successive spectral lines is clearly larger for a thinner QW, therefore we conclude that the observed spectral series should be related with some transversal size quantization. It is seen that optical spectra contain several lines marked $L_1$-$L_4$ that are less intensive than their neighbours; these lines demonstrate also different temperature and magnetic field dependencies as will be discussed later on. For further analysis we select spectral lines with similar amplitudes (in Fig.1 they are marked by arrows looking upward) and try to associate their positions with some translational quantization indices N. Instead of fixing some translational mass $M$ as in ~\cite{Kochereshko} and extracting corresponding N from the well-known relation $E_N=E_0+\frac{\hbar^2}{2M}{(\frac{N\pi}{d}})^2$, the lowest line energy was prescribed with some index $N_0$ and all higher energies got indices $N_0+1$, $N_0+2$ and so on ~\cite{Note0}. There should be only one series of indices that provides a straight line in coordinates Energy vs. $N^2$ ~\cite{Note1}. Figure 2 demonstrates how the experimental points fit into a linear dependence for several choices of $N_0$. The most appropriate values of $N_0$ for both samples are outlined by rectangles and they give a perfect coincidence with a size quantization model with the same value of $E_0=1.5148$ eV. It is these numbers that mark the spectral lines below the arrows in Fig.1 that are obtained from this simple fit. 

The linear slopes of Fig.2 allow us to determine the values of translational mass $M$. Surprisingly, they equal 0.084$m_e$ and 0.092$m_e$ for QW of 250 nm and 150 nm in width respectively. This is the main finding of this paper: the extracted masses cannot be related with any of heavy-hole or light-hole excitons possible for GaAs material. Instead, we must conclude that the optical spectra are much better described by quantization of a light hole only because its measured mass of (0.078$\pm$0.002)$m_e$ in a bulk state ~\cite{Gubarev} is very close to our experimental results for $M$ ~\cite{Note11}. 

To support this model further, a magnetic field was applied in the growth direction of the sample to investigate its effect on optical spectra. Figure 3 shows the measured splittings of PL spectral lines with $N$=2 and $N$=3 from the 150 nm QW in a magnetic field up to 2 Tesla. These splittings deviate slightly from a linear dependence at high B-fields and can be approximately described by effective g-factors with absolute values of 2.7 and 5.3 for the $N$=2 and $N$=3 lines correspondingly. The obtained values are much higher than either electron g-factor $g_e$=-0.4, heavy hole g-factor $g_{hh}$=0.3 or their combination $g_e-3g_{hh}$=-1.3 which describes the Zeeman effect for $\sigma^{\pm}$ optical transitions for a heavy-hole exciton. In turn, the light holes are known to have anomalously large g-factor ~\cite{Gubarev} because of the spin-orbit interaction and k$\cdot$p mixing of the electron bands. Moreover, a recent theoretical study ~\cite{Glazov} predicts a light hole g-factor in a QW to be dependent on both a QW width and a quantized subband involved. For higher subband indices this parameter should increase due to larger values of $K_z$ in accordance with our measurements, however the full theoretical description of that phenomenon is still lacking.

Observation of light holes signatures in the optical spectra poses a question which should be discussed: why do heavy holes reveal no clear picture of quantization as opposed to light holes?

It is well known that a QW valence band hole is characterized by a complex dispersion relation due to a size quantization. As pointed out in ~\cite{Dyakonov}, the physical reason for that is a nonzero probability for a heavy hole to be transformed into a light hole (and vice versa) in the course of the reflection from the QW wall. The resulting hole wave function is a linear combination of wave functions for light and heavy holes in their bulk states and the in-plane dispersion relations for all subbands turn out to be strongly renormalized. The most prominent change takes place for heavy holes where the calculated effective masses deviate drastically from the bulk value and even become negative for some subband indices $n$ ~\cite{Dyakonov}. This is the first possible reason why some heavy-hole subbands can be optically inactive. On the contrary, all light-hole subbands have a monotonically decreasing and relatively weak dependence of an in-plane mass on the subband index. Another reason might be an essential difference in scattering lengths for both types of holes. Size quantization in a QW requires that a particle preserves its coherence between successive reflections from the QW barriers. The lighter particle is expected to have a longer scattering length and therefore can reveal quantum properties at large scales. In our samples the QWs may be "`thin enough"' for light holes while heavy holes experience one or more scatterings at distances of several multiples of 150-250 nm.

One more consequence of the complex valence band structure is the allowance of the "`forbidden"' optical transitions between electrons and holes with different $N$ ~\cite{Dyakonov}. This explains the fact that the discussed spectral features correspond to the transitions between the 1st electron subband and the N-th light hole subbands otherwise optically inactive for all $N>$1. Spectral lines $L_1$-$L_4$ in Fig.1 can involve optical transitions to higher electron subbands or even might be attributed to some heavy-hole QW states, however these features were not investigated so far in detail. The absence of prominent optical transitions from higher electron subbands is tentatively explained by residual p-doping of our nominally undoped structures which results to a short scattering length for photoexcited electrons as compared with photoexcited light holes.

Bearing in mind possible contribution of excitonic effects into a quantization picture for our QWs, we also studied temperature-dependent reflectance spectra. The binding energy of a heavy-hole exciton in GaAs is approximately $Ry$=5 meV which corresponds to the dissociation temperature of some 60 K. This value is even smaller for a light-hole exciton. Temperature increase up to the dissociation point and above it leads to a broadening of the hydrogen-like exciton energy levels due to a thermal escape into continuum states. The thermal broadening can be simply estimated to be equal to $\Gamma(T)=\gamma_0 e^{(-\frac{Ry}{T})}$ , where $\gamma_0$=5$\div$10 meV as obtained from early experiments and theoretical considerations ~\cite{Lifetime}. An image of temperature-dependent reflectivity spectra from a 250 nm QW is shown in Figure 4. The bandgap energy change results to a red-shift of the observed optical transitions when the temperature is increased from 1.5 to 77 K. However the most prominent spectral features remain nearly unchanged. Figure 5 compares two reflectivity spectra recorded at 1.5 and 77 K with the low-temperature data red-shifted by 7 meV to compensate for a band gap change. Well-resolved spectral lines are observed again with linewidths of approximately 0.5 meV and an	unambiguous association between the two series is possible. At 77 K any exciton-related spectral feature should have a thermal level broadening of more than 1 meV, therefore we conclude again that our experimental findings cannot be explained in terms of exciton quantization.

To summarize, we have investigated an optical response of ultra-wide GaAs quantum wells by means of photoluminescence and reflectance spectroscopy. It is shown that a picture of light holes quantization in a QW describes quite well our experimental results. From the analysis of temperature-dependent spectra we exclude an exciton quantization from possible models explaining the findings.

Support from the RFBR is greatly acknowledged. 

\medskip

\bfseries Figure Captions: \normalfont

\bfseries FIGURE 1 \normalfont

Photoluminescence and reflectivity spectra from 250 nm (top) and 150 nm (bottom) ultra-wide quantum wells detected at 1.5 K. The most prominent spectral lines are marked with a corresponding translational quantization index N.

\bfseries FIGURE 2 \normalfont

Procedure of quantization indices matching for spectral lines from Fig.1. The lowest optical transition from a spectral series is associated with a successively increasing index. The only designation delivers a perfectly straight line in coordinates chosen for our experimental points. From a slope of the line the effective mass $M$ of a quantized particle is extracted.

\bfseries FIGURE 3 \normalfont

Zeeman splittings of optical transitions with $N$=2 and $N$=3 from the PL spectra of 150 nm QW and the corresponding effective g-factors. 

\bfseries FIGURE 4 \normalfont

Reflectivity spectra from a QW 250 nm thick, recorded at temperatures from 1.5 K to 77 K. All spectral lines experience a red-shift in energy due to a band gap energy decreasing, however most lines detected at the lowest temperatures are also observed at 77 K. Dotted line traces the optical transition corresponding to $N=9$ from Fig.1, as an example.

\bfseries FIGURE 5 \normalfont

Comparison of reflectivity spectra from a QW 250 nm thick at two different temperatures of 1.5 and 77 K, the former red-shifted by 7 meV to match energies of the lowest optical transitions. The main series of spectral lines is still observed at higher temperatures while the lines $L_1$ and $L_2$ from Fig.1 has disappeared.

\end{document}